# Quantifying Chemical Structure and Atomic Energies in Amorphous Silicon Networks

Noam Bernstein,[a] Bishal Bhattarai,[b] Gábor Csányi,[c] David A. Drabold,[b]
Stephen R. Elliott,[d] and Volker L. Deringer[c,d],*

[a]*Center for Materials Physics and Technology, U.S. Naval Research Laboratory, Washington, DC 20375, United States*

[b]*Department of Physics and Astronomy, Ohio University, Athens, Ohio 45701, United States*

[c]*Department of Engineering, University of Cambridge, Cambridge CB2 1PZ, UK*

[d]*Department of Chemistry, University of Cambridge, Cambridge CB2 1EW, UK*

\* Corresponding author. E-mail: vld24@cam.ac.uk

**Abstract:** Amorphous materials are coming within reach of realistic computer simulations, but new approaches are needed to fully understand their intricate atomic structures. Here, we show how machine-learning (ML)-based techniques can give new, quantitative chemical insight into the atomic-scale structure of amorphous silicon (*a*-Si). Based on a similarity function ("kernel"), we define a structural metric that unifies the description of nearest- and next-nearest-neighbor environments in the amorphous state. We apply this to an ensemble of *a*-Si networks, generated in melt–quench simulations with an ML-based interatomic potential, in which we tailor the degree of ordering by varying the quench rates down to $10^{10}$ K/s (leading to a structural model that is lower in energy than the established WWW network). We then show how "machine-learned" atomic energies permit a chemical interpretation, associating coordination defects in *a*-Si with distinct energetic stability regions. The approach is straightforward and inexpensive to apply to arbitrary structural models, and it is therefore expected to have more general significance for developing a quantitative understanding of the amorphous state.

The structure of amorphous silicon (*a*-Si) is widely approximated as a continuous random network with tetrahedral coordination,[1] but its details are much more intricate: defective environments, such as threefold-bonded "dangling bonds", as well as the degree of medium-range order, have been discussed.[2] Together with experimental probes,[3] atomistic computer simulations have been giving useful insight into *a*-Si for decades,[4] and large-scale simulation models now contain up to hundreds of thousands of atoms.[5] With the recent emergence of linear-scaling machine-learning (ML)-based interatomic potentials, reaching accuracy levels close to quantum mechanics,[6] materials modeling is promising to become even more realistic – especially in describing amorphous solids,[7] as recently shown for *a*-Si.[8]

Still, there remains the more fundamental challenge not only to describe amorphous structures but to truly understand them. Simple criteria are widely used, including the counting of atomic coordination numbers (here denoted $N$) and the "tetrahedral-likeness" which both measure short-range order (SRO),[9] or ring statistics as a proxy for medium-range order (MRO).[10] However, a unified approach for both length scales has been missing so far. And more critically, these structurally-based indicators cannot give information about the energetic stability of individual atomic environments.

In this contribution, we describe a general way to quantify both local structures and local energies of all individual atoms in amorphous solids. Our object of study is an ensemble of *a*-Si networks that we created in parallel ML-driven molecular-dynamics (MD) simulations, in which 512-atom liquid Si models were cooled to solidify into *a*-Si (Figure 1a). Slower cooling yields more ordered networks:[8] hence, changing the cooling rate allows us to tailor the degree of order in the structures and to probe its influence on properties. Remarkably, the most ordered structures we obtained ($10^{11}$ and $10^{10}$ K/s), albeit still containing ≈1% defects, are energetically more favorable by 0.02 eV/at. (DFT-PBE) than a fully tetrahedral Wooten–Winer–Weaire (WWW) model,[1] the latter so far being a "gold standard" for *a*-Si (Supporting Information).

An established indicator for SRO in *a*-Si is to measure how similar the atomic environments are to ideal tetrahedra, as probed *via* the bond angles.[9] In our set of structures, this indicator increases as expected with slower quenching (increasing ordering) and then converges, such that the median results for the $10^{11}$ and $10^{10}$ K/s structures are very similar (Figure 1b). There is large scatter across different atoms in the cells (seen from the first and third quartiles), again as expected in the amorphous state. We also look at MRO *via* the count of six-membered rings (Figure 1c) – which is 100% in diamond-type *c*-Si, where all atoms are in cyclohexane-like rings, but smaller in *a*-Si due to the presence of disorder. We stress that the indicator in Figure 1b cannot fully quantify MRO, as it focuses on angles alone; the ring count in Figure 1c, on the other hand, does not give information about SRO.

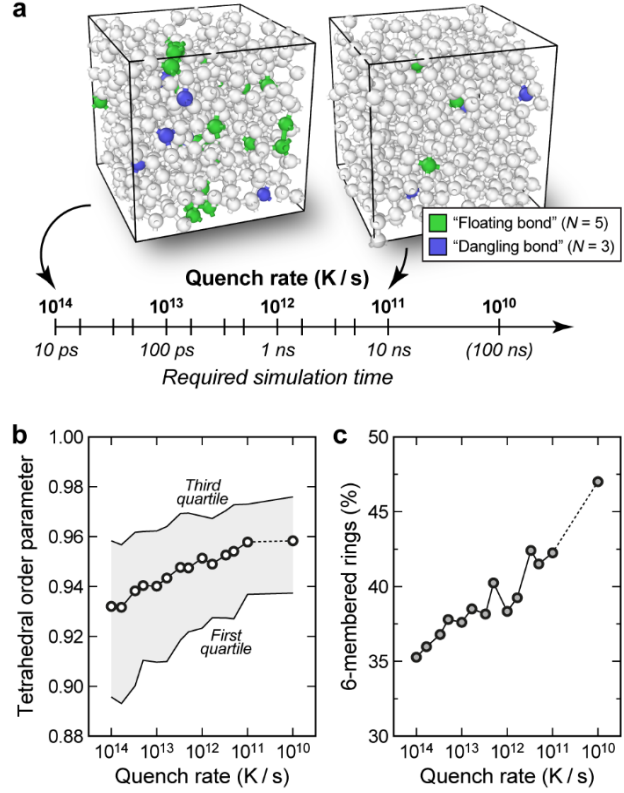

**Figure 1.** Progressively ordered *a*-Si networks from melt–quench simulations with an ML-based interatomic potential of quantum-mechanical quality. (a) The scale of cooling rates and associated required simulation times (one picosecond requires 1,000 MD time steps). Each tick corresponds to one independent MD simulation. Between $10^{14}$ and $10^{11}$ K/s, we cooled at the respective constant rate; for the much more demanding $10^{10}$ K/s simulation, we varied the rate during the run (Supporting Information). Two simulation cells are shown as examples, and coordination defects are highlighted by coloring (over-coordinated "floating-bond" environments in green; under-coordinated "dangling-bond" environments in blue). (b) Increasing structural order in these systems, quantified using an established order parameter that returns unity for ideal tetrahedral environments.[9] (c) Increasing MRO, assessed by counting 6-membered rings.[10]

To progress further, we now turn to the Smooth Overlap of Atomic Positions (SOAP) kernel,[11] a mathematical approach that has been used with success to fit ML potentials[12] and to analyze structures.[13] We expand the smoothed neighbor density $\rho_i(\mathbf{r})$ around the *i*-th atom (Figure 2a) into an atom-centered basis of radial basis functions $R_n$ and spherical harmonics $Y_{lm}$,[11]

$$\rho_i(\mathbf{r}) = \sum_{nlm} c_{nlm}^{(i)} R_n(r) Y_{lm}(\hat{\mathbf{r}}),$$

similarly to how electronic wave functions are expanded in quantum chemistry. By combining the coefficients $c_{nlm}$ into the rotationally invariant power spectrum, **p**, the similarity of two environments $(i, j)$ is obtained as a simple dot product: [11]

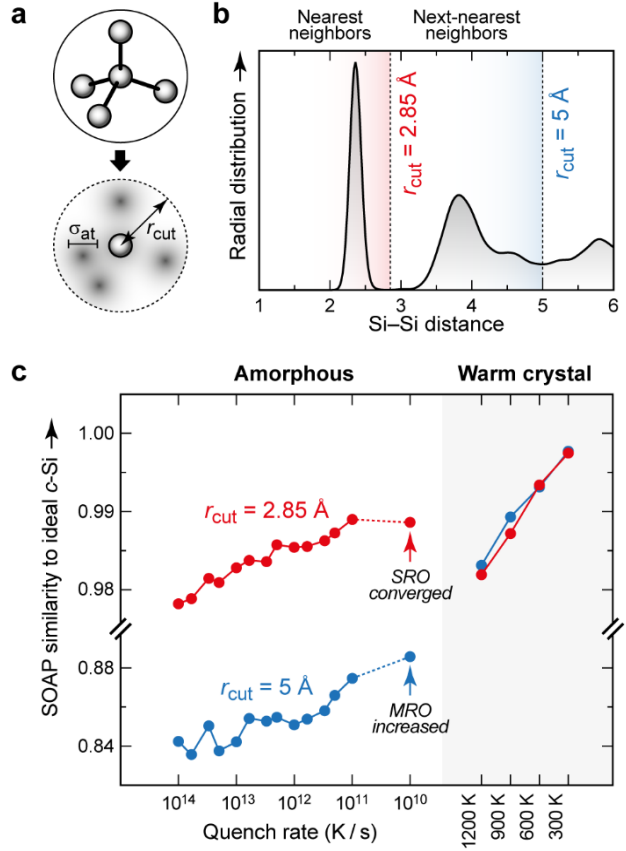

**Figure 2.** Quantifying local structure in a series of progressively ordered *a*-Si networks. (a) The SOAP approach translates an atomic environment into a smoothed neighbor density within a given cutoff, $r_{\text{cut}}$, whereby the parameter $\sigma_{\text{at}}$ controls the smoothness.[11] The resulting density can then be compared for a pair of environments using the SOAP kernel, $K_{ij}$. (b) Radial distribution function (RDF) for the $10^{10}$ K/s structure, illustrating the choice of two different cutoff radii that reach up to nearest-neighbor (NN) and next-nearest neighbor (NNN) environments, respectively. (c) SOAP kernel evaluated for NNs (*red*) and NNNs (*blue*) in the different *a*-Si networks, using the parameters in Table 1, quantifying similarity to diamond-type *c*-Si. Median values over all atoms in the cells are given for each system. For calibration, we also include a 512-atom *c*-Si cell which was heated to increasingly higher temperatures in GAP-MD simulations. The parameters used are listed in Table 1.

$$p^{(i)}_{nn'l} \propto \sum_m c^{(i)*}_{nlm} c^{(i)}_{n'lm} \quad \rightarrow \quad K_{ij} = \left[ \mathbf{p}^{(i)} \cdot \mathbf{p}^{(j)} \right]^\zeta \text{ (here, } \zeta=4\text{)},$$

with the kernel function $K_{ij}$ being 1 for environments that are identical (up to symmetry operations) and 0 for environments that are maximally dissimilar. In this work, we use a *pair* of SOAP kernels, one ranging up to nearest-neighbor (NN) environments and one extending further to include also the next-nearest neighbors (NNN; Figure 2b). We use these to compare each atom in our *a*-Si structures to *c*-Si and report the median value. Figure 2c shows that the SRO in our *a*-Si networks converges with slower quenching, while the MRO increases further all the way to $10^{10}$ K/s. This is consistent with our finding for the ring sizes but is now expressed in one and the same mathematical framework.

**Table 1.** SOAP parameters[11] for the pair of kernels defined in this work, and results for atomic sites in crystalline Si allotropes as obtained from both.

|  |  | NN kernel | NNN kernel |
|---|---|---|---|
| Basis set size ($n_{max}$, $l_{max}$) | | (16, 16) | (16, 16) |
| Cutoff radius $r_{cut}$ (Å) | | 2.85 | 5.00 |
| Transition width $r_\Delta$ (Å) | | 0.30 | 0.60 |
| Smoothness $\sigma_{at}$ (Å) | | 0.30 | 0.60 |
| Similarity to diamond-type *c*-Si in… | | | |
| [lonsdaleite] | Si1 ≡ Si2 (2*b*) | 1.000 | 0.974 |
| *oS*24[14] | Si1 (8*f*) | 0.980 | 0.792 |
|  | Si2 (8*f*) | 0.994 | 0.908 |
|  | Si3 (8*f*) | 0.987 | 0.747 |
| [β-tin] | Si (4*a*) | 0.723 | 0.342 |

While SOAP provides a quantitative similarity measure on a scale from 0 to 1, the absolute value depends on many parameters, including the "fuzziness" of the Gaussians that are placed on atomic neighbors (*cf.* Figure 2a).[11,12] This is a challenge here, as we wish to combine two different SOAPs: that for the NN shell making a sharp distinction between environments, and that for the NNN shell being more tolerant to small structural changes. Therefore, we calibrated the SOAP "fuzziness" ($\sigma_{at}$) using *c*-Si at *T* > 0 K as reference, by requiring that NN and NNN SOAP results are similar in an ordered network with only thermal fluctuations.

The simplicity and power of this approach can also be shown by applying it to other crystalline allotropes of Si (Table 1). In the lonsdaleite- ("hexagonal diamond-") type form,[15] the NN environments are as in the cubic diamond type (namely, ideal tetrahedra, with a similarity of 1.000) – but the NNN environments differ, due to rings being in "boat" rather than "chair" conformations, and hence the similarity to *c*-Si drops to 0.974. We next look at an open-framework Si allotrope, *oS*24, which was synthesized from Na$_4$Si$_{24}$ by sodium de-intercalation.[14] In *oS*24, the atoms are tetrahedrally coordinated too, but with strong local distortions, and so the resulting NN values are comparable to those in *a*-Si; the NNN values drops further, because the open framework structure is remarkably different from *c*-Si. Finally, for β-tin-type Si with "4+2" coordination, the NN environments are clearly dissimilar from those in *c*-Si, and the NNNs even more so (Table 1).

We now ask for the energies of individual atoms, a crucial piece of information that cannot easily be obtained from DFT computations (which yield the total energy for the entire cell). In contrast, atomic energies are directly included in many ML-based interatomic potentials by construction.[6,16,17] In the Gaussian Approximation Potential (GAP) framework we use here, the total energy is a sum of "machine-learned" atomic energies, and the general form for the latter is[16]

$$\varepsilon_i = \sum_j \alpha_j K_{ij},$$

where the sum runs over a set of reference environments in the training database (index $j$) and two environments are compared using the kernel function $K_{ij}$. Recall that when using SOAP, the latter is evaluated as a simple dot product: the evaluation of structural similarity and of local energies in this work therefore uses the same mathematical foundation.[11] We recently showed that these "machine-learned" atomic energies can carry chemical meaning: namely, for β-rhombohedral boron, where sites with partial vacancy formation were shown to be associated with high (*i.e.*, unfavorable) GAP atomic energies.[12d] In the present work, we now transfer this concept to the amorphous state, where there is an even more dire need for information about atomically resolved stabilities. We note that such analyses are in principle possible with empirical interatomic potentials,[18] but can be limited by the parametric shape of the potential. In contrast, our approach depends only on the input data, is readily generalized, and combines accurate DFT input data with a high-fidelity ML fit whose uncertainty can be quantified[12e] to be in the meV range (Supporting Information).

Figure 3a shows that indeed, "machine-learned" atomic energies recover the stability trends intuitively expected for *a*-Si, the interpretation being qualitative for now. A dangling-bond defect (*red*) shows high local energy, whereas the two tetrahedral-like central atoms (*white*, *blue*) are more energetically favorable, depending on how strongly they are distorted. Histograms of these data, collected for a disordered, fast-quenched structure (Figure 3b) and a more ordered, slowly-quenched structure (Figure 3c) reveal that the center of gravity for the more ordered *a*-Si network does coincide with the experimentally determined stability.[19]

A 2D plot of both quantities, SOAP-based diamond-similarity (*x*-axis) and local energy (*y*-axis), is perhaps most revealing. This plot is given in Figure 3d. Distinct energy regions, at around +0.4 and +0.6 eV greater than that of c-Si, are observed for floating-bond ($N = 5$) and dangling-bond ($N = 3$) environments. The floating bonds show a wide structural variation within the NN shell, indicated by a large spread over the SOAP (*x*-axis) coordinate, whereas the dangling bonds clearly peak around an NN similarity value of 0.92. In this, the dangling-bond atoms are structurally closer to *c*-Si than are the "4+2"-coordinated atoms in the [β-tin] allotrope (Table 1), but they are considerably less "diamond-like" than the median result for any of our *a*-Si structures (Figure 2c), which are dominated by fourfold-bonded atoms. Looking at the plot in Figure 3d again, it appears that the data-points for dangling bonds ($N = 3$) lie in the tail of a continuation of the GAP energy–SOAP similarity plot for tetrahedral-like environments ($N = 4$); this is not the case for floating bonds ($N = 5$). Finally, we note that the GAP energies for $N = 4$ atoms reach up to high values: their median value is +0.14 eV, but the 98th percentile is at +0.42 eV, and therefore the remaining 2% fourfold-bonded atoms have energies that are higher than the median result for N = 5 defects (+0.42 eV). This explains how our $10^{10}$ K/s structure, albeit having defects, can be lower in energy than the defect-free WWW model.[1]

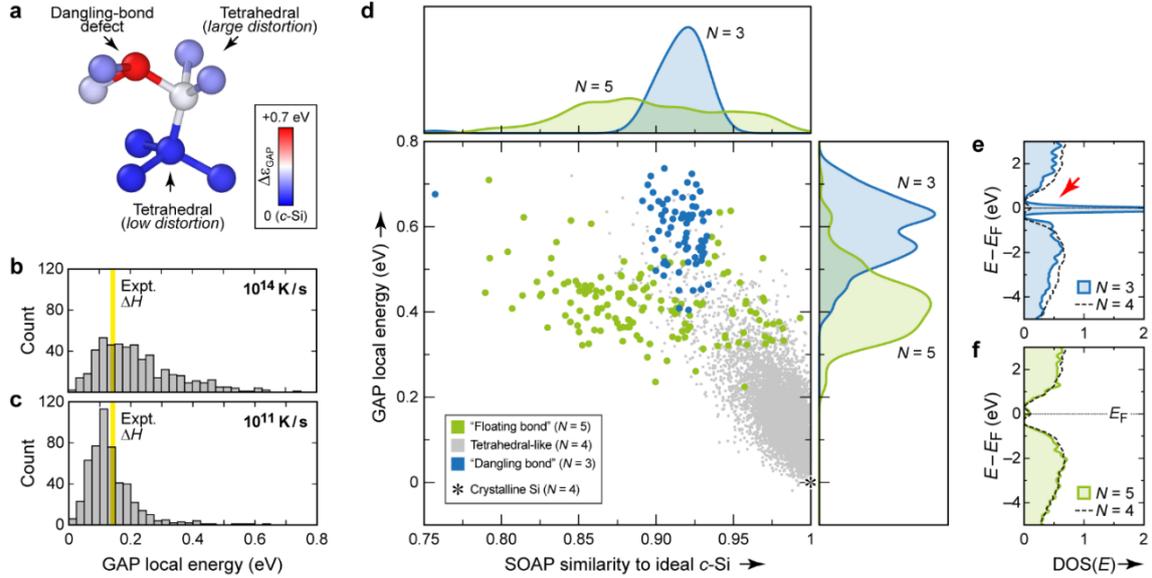

**Figure 3.** "Machine-learned" atomic energies in *a*-Si. (a) Example structural fragment, chosen to represent a dangling-bond defect with high energy (*red*), a distorted tetrahedral environment with intermediate energy (*white*), and a more favorable tetrahedral environment showing only low distortion (*blue*). Atoms are color-coded according to their GAP atomic energy, $\varepsilon_i$, given relative to that in diamond-type *c*-Si. (b) Histogram of atomic energies in the structure quenched at $10^{14}$ K/s. The experimental enthalpy of relaxed *a*-Si (from Ref. [19], also relative to *c*-Si) is indicated by yellow shading. (c) Same but for the more ordered $10^{11}$ K/s structure. (d) 2D plot revealing the connection between structural order (NN similarity to diamond-like *c*-Si; horizontal axis) and GAP local energy for the individual atoms (vertical axis). Results are collected for all 14 systems, *i.e.*, for all quench rates from $10^{14}$ to $10^{10}$ K/s (*cf.* Figure 1). Kernel density estimates ("smoothed histograms") are given for projections on both axes. (e,f) Local electronic DOS for a $10^{11}$ K/s quenched structure from Ref. [8], illustrating the very different electronic fingerprints of three- and fivefold-bonded coordination defects. DOS plots are normalized per atom; for comparison, the average local DOS for all fourfold bonded atoms in the same structure is given by dashed lines. The red arrow in (e) highlights the mid-gap states associated with dangling-bond defects.

The higher GAP atomic energy (*i.e.*, larger instability) of dangling compared to floating bonds is not only in line with previous theories,[2a,18] but it can also be corroborated by the electronic densities of states (DOS). For the energetically unfavorable dangling bonds ($N = 3$), a large peak at the Fermi level, within the bandgap, is seen from an atomically resolved projection of the DOS (arrow in Figure 3e). In contrast, the more favorable floating bonds ($N = 5$) show no such pronounced mid-gap contributions (Figure 3f), especially when compared to a properly normalized projection on the tetrahedral-like environments (dashed line).

In conclusion, we have here described a general framework for quantifying not only local structure but also local atomic energies in amorphous silicon networks. A combined plot of both quantities (Figure 3d) yields an instructive and chemically interpretable overview of the atomically resolved properties of *a*-Si. The approach is expected to be more general, and to have wider significance: it could be applied to other tetrahedral networks, such as the homologous germanium, to amorphous silicates or metal–organic frameworks,[20] where similarly tetrahedral building units are found, or even to crystalline, amorphous, and liquid water.[9b,21]

**Acknowledgements**

N.B. acknowledges support from the Office of Naval Research through the U. S. Naval Research Laboratory's core basic research program, and computer time from the DoD High Performance Computing Modernization Program Office at the Air Force Research Laboratory DoD Supercomputing Resource Center. D.A.D. thanks the US NSF for support from grant DMR 1506836. V.L.D. acknowledges a Leverhulme Early Career Fellowship and support from the Isaac Newton Trust. This work used the ARCHER UK National Supercomputing Service via EPSRC Grant EP/P022596/1.

**References**

[1]   a) F. Wooten, K. Winer, D. Weaire, *Phys. Rev. Lett.* **1985**, *54*, 1392; b) B. R. Djordjević, M. F. Thorpe, F. Wooten, *Phys. Rev. B* **1995**, *52*, 5685.

[2]   a) S. T. Pantelides, *Phys. Rev. Lett.* **1986**, *57*, 2979; b) R. Xie et al., *Proc. Natl. Acad. Sci., U. S. A.* **2013**, *110*, 13250.

[3]   a) W.-L. Shao, J. Shinar, B. C. Gerstein, F. Li, J. S. Lannin, *Phys. Rev. B* **1990**, *41*, 9491; b) P. Roura, J. Farjas, P. Roca i Cabarrocas, *J. Appl. Phys.* **2008**, *104*, 073521; c) A. M. R. De Graff, M. F. Thorpe, *Acta Crystallogr., Sect. A* **2010**, *66*, 22.

[4]   Examples: a) D. A. Drabold, P. A. Fedders, O. F. Sankey, J. D. Dow, *Phys. Rev. B* **1990**, *42*, 5135; b) I. Štich, R. Car, M. Parrinello, *Phys. Rev. B* **1991**, *44*, 11092; c) N. Bernstein, J. L. Feldman, M. Fornari, *Phys. Rev. B* **2006**, *74*, 205202.

[5]   R. Atta-Fynn, P. Biswas, *J. Chem. Phys.* **2018**, *148*, 204503.

[6]   J. Behler, *Angew. Chem. Int. Ed.* **2017**, *56*, 12828, and references therein.

[7]   a) G. C. Sosso, G. Miceli, S. Caravati, F. Giberti, J. Behler, M. Bernasconi, *J. Phys. Chem. Lett.* **2013**, *4*, 4241; b) M. A. Caro, V. L. Deringer, J. Koskinen, T. Laurila, G. Csányi, *Phys. Rev. Lett.* **2018**, *120*, 166101; c) V. L. Deringer, C. Merlet, Y. Hu, T. H. Lee, J. A. Kattirtzi, O. Pecher, G. Csányi, S. R. Elliott, C. P. Grey, *Chem. Commun.* **2018**, *54*, 5988; d) N. Artrith, A. Urban, G. Ceder, *J. Chem. Phys.* **2018**, *148*, 241711; e) B. Onat, E. D. Cubuk, B. D. Malone, E. Kaxiras, *Phys. Rev. B* **2018**, *97*, 094106; f) V. Lacivita, N. Artrith, G. Ceder, *Chem. Mater.* **2018**, *30*, 7077.

[8]   The computational protocol has been introduced and validated against experimental probes in our previous work: V. L. Deringer, N. Bernstein, A. P. Bartók, R. N. Kerber, M. J. Cliffe, L. E. Marbella, C. P. Grey, S. R. Elliott, G. Csányi, *J. Phys. Chem. Lett.* **2018**, *9*, 2879. In the present work, however, we generated an entirely new set of structures to sample more finely spaced quench rates (Figure 1a), and we expand the scope of such simulations to a slowest rate of $10^{10}$ K/s.

[9]   a) P.-L. Chau, A. J. Hardwick, *Mol. Phys.* **1998**, *93*, 511; b) J. R. Errington, P. G. Debenedetti, *Nature* **2001**, *409*, 318.

[10]  D. S. Franzblau, *Phys. Rev. B* **1991**, *44*, 4925.

[11]  A. P. Bartók, R. Kondor, G. Csányi, *Phys. Rev. B* **2013**, *87*, 184115.

[12]  Examples: a) A. P. Bartók, S. De, C. Poelking, N. Bernstein, J. R. Kermode, G. Csányi, M. Ceriotti, *Sci. Adv.* **2017**, *3*, e1701816; b) V. L. Deringer, G. Csányi, *Phys. Rev. B*

 **2017**, *95*, 094203; c) D. Dragoni, T. D. Daff, G. Csányi, N. Marzari, *Phys. Rev. Mater.* **2018**, *2*, 013808; d) V. L. Deringer, C. J. Pickard, G. Csányi, *Phys. Rev. Lett.* **2018**, *120*, 156001; e) A. P. Bartók, J. R. Kermode, N. Bernstein, G. Csányi, *Phys. Rev. X*, in press [preprint at arXiv:1805.01568].

[13] a) S. De, A. P. Bartók, G. Csányi, M. Ceriotti, *Phys. Chem. Chem. Phys.* **2016**, *18*, 13754; b) J. Mavračić, F. C. Mocanu, V. L. Deringer, G. Csányi, S. R. Elliott, *J. Phys. Chem. Lett.* **2018**, *9*, 2985; c) M. O. J. Jäger, E. V. Morooka, F. F. Canova, L. Himanen, A. S. Foster, *npj Comput. Mater.* **2018**, *4*, 37; d) M. A. Caro, A. Aarva, V. L. Deringer, G. Csányi, T. Laurila, *Chem. Mater.* **2018**, *30*, 7446.

[14] D. Y. Kim, S. Stefanoski, O. O. Kurakevych, T. A. Strobel, *Nat. Mater.* **2015**, *14*, 169.

[15] R. H. Wentorf, Jr., J. S. Kasper, *Science* **1962**, *139*, 338.

[16] A. P. Bartók, M. C. Payne, R. Kondor, G. Csányi, *Phys. Rev. Lett.* **2010**, *104*, 136403.

[17] The usefulness of ML-based local energies for global structure optimization was also shown very recently: a) T. L. Jacobsen, M. S. Jørgensen, B. Hammer, *Phys. Rev. Lett.* **2018**, *120*, 026102; b) X. Chen, M. S. Jørgensen, J. Li, B. Hammer, *J. Chem. Theory Comput.* **2018**, *14*, 3933.

[18] P. C. Kelires, J. Tersoff, *Phys. Rev. Lett.* **1988**, *61*, 562.

[19] S. Roorda, W. C. Sinke, J. M. Poate, D. C. Jacobson, S. Dierker, B. S. Dennis, D. J. Eaglesham, F. Spaepen, P. Fuoss, *Phys. Rev. B* **1991**, *44*, 3702.

[20] T. D. Bennett, A. K. Cheetham, *Acc. Chem. Res.* **2014**, *47*, 1555.

[21] Large-scale extended water networks are also becoming accessible to ML-based modeling. A recent study successfully described the melting point of water (to within 1 K): T. Morawietz, A. Singraber, C. Dellago, J. Behler, *Proc. Natl. Acad. Sci., U. S. A.* **2016**, *113*, 8368.

Supporting Information for

# Quantifying Chemical Structure and Atomic Energies in Amorphous Silicon Networks

Noam Bernstein, Bishal Bhattarai, Gábor Csányi, David A. Drabold,
Stephen R. Elliott, and Volker L. Deringer*

**This Supporting Information document contains:**

———

*E-mail: vld24@cam.ac.uk

# Computational details

*Generation of a-Si networks*

Constant-pressure MD simulations (NPT ensemble), driven by a general-purpose GAP for Si (with the unique identifier `GAP_2017_6_17_60_4_3_56_165`),[S1] were carried out using LAMMPS.[S2–5] The system size was 512 atoms per cell, and the time step was 1 fs. After initial mixing at 1,800 K and keeping the system in the liquid state at 1,500 K, a quench to 500 K was performed with a given quench rate, and a final optimization was done using a conjugate-gradient (CG) minimizer. The computational protocol has been introduced and validated in a previous, more technical work;[S6] this includes comparison to the experimental structure factor and to experimentally measured $^{29}$Si NMR chemical shifts (see [S6] and references therein).

In the present work, an entirely new set of structures was generated to sample more finely spaced quench rates (*cf.* Figure 1a). Importantly, we also expanded the scope of the simulations to a slowest rate of $10^{10}$ K/s, yielding a model with very high structural ordering (Figures 1c and 2c). A constant-rate quench at this rate would require 100 ns of simulation time ($10^8$ time steps, or several months of real time). We therefore use a variable-rate quenching scheme which was introduced and validated in [S6], in which the slowest quench rate (here, $10^{10}$ K/s) is only applied in the temperature region where it is found to be truly needed (viz. between 1,250 and 1,050 K), and a faster rate of $10^{13}$ K/s is applied elsewhere. All structural models obtained and analyzed in this work are provided as separate Supporting Information.

*Local energies from GAP*

Among the key findings of this work is that local energies, as obtained from an ML-based interatomic potential by construction, permit chemical interpretation. We review the most important aspects here for convenience; the interested reader is referred to the original literature for a more detailed derivation.[S7–9] The local energy of the *i*-th atom, $\varepsilon_i$, is obtained in the Gaussian Approximation Potential (GAP) framework as follows:[S7]

$$\varepsilon_i = \sum_j \alpha_j K(\mathbf{q}_i, \mathbf{q}_j) \equiv \sum_j \alpha_j K_{ij}$$

(where the latter is just a shorthand notation). The sum runs over a set of reference configurations in the training database (index *j*), each described by a general descriptor vector denoted "**q**", of which each is compared to the atom in question using the kernel function *K*. In the case of SOAP, this kernel is simply a properly normalized dot product, raised to a small integer power for better distinction between environments (here, $\zeta = 4$).[S8]

Depending on the physical nature of the system, more complex formulations may be used, but the general idea stays the same. For example, multiple descriptors (label *d*) can be combined by forming a linear combination with appropriate scaling factors $\delta^{(d)}$:[S9–11]

$$\varepsilon_i = \sum_d \delta^{(d)} \left( \sum_j \alpha_j^{(d)} K^{(d)}\left(\mathbf{q}_i^{(d)}, \mathbf{q}_j^{(d)}\right) \right).$$

Again, the result is one atomic energy value, $\varepsilon_i$. In the general-purpose silicon GAP we use, finally, the repulsive interaction between atoms at small distances is covered by a parametric two-body term ("core potential"), and the SOAP kernel is then used in the second term:[S1]

$$\varepsilon_i = \sum_j V_{ij}(r_{ij}) + \sum_j \alpha_j K(\mathbf{q}_i, \mathbf{q}_j).$$

For the sake of brevity, we have omitted the leading two-body potential term from the presentation in the main text. All these formulations lead to well-defined atomic energies $\varepsilon_i$, and we note that similar approaches can be followed with other ML-based interatomic potentials such as artificial neural networks.[S12–14]

The full dataset of atomic energies is provided in extended `XYZ` format (to be directly read into programs such as `ASE` or `quippy`) and as a separate `csv` file. For convenience, in **Table S1**, we provide percentile values.

**Table S1:** Percentile values of the distribution of local energies (the 98-th percentile value means that 98 percent of the atoms have a local GAP energy below this value, *etc.*).

| Percentile | $N = 3$ only | $N = 4$ only | $N = 5$ only | All atoms |
|---|---|---|---|---|
| **100 (Maximum)** | **0.737** | **0.720** | **0.710** | **0.737** |
| 98 | 0.722 | 0.418 | 0.623 | 0.486 |
| 90 | 0.681 | 0.281 | 0.521 | 0.306 |
| 75 | 0.645 | 0.206 | 0.463 | 0.214 |
| **50 (Median)** | **0.598** | **0.144** | **0.417** | **0.148** |
| 25 | 0.544 | 0.101 | 0.364 | 0.103 |
| 10 | 0.494 | 0.070 | 0.332 | 0.071 |
| 2 | 0.427 | 0.037 | 0.283 | 0.038 |
| **0 (Minimum)** | **0.405** | **–0.019**[a] | **0.224** | **–0.019**[a] |

[a]The lowest value is slightly more negative than that of an atom in the optimized diamond structure, but the difference is within the maximum expected error (page S7).

*Electronic structure*

The local DOS plots shown in Figure 3e–f were obtained using an analytic projection scheme, as implemented in `LOBSTER`,[S15,16] ensuring a quantitative transfer from a Γ-point VASP-PBE computation into a local basis of Slater-type orbitals (Si 3s 3p). The electronic wavefunction and the total DOS had already been computed in our previous work;[S6] here, this allows us to conveniently illustrate the effect of local environments. Further studies are currently underway to study more closely the electronic structure of the structural models generated here; these will be reported elsewhere.

*Implementation and visualization*

The local energies, including the predicted error, can be obtained with the freely available `QUIP/quippy` software (all required software, including the GAP prediction code, can be obtained at http://www.libatoms.org). The following `python` script yields this information:

**Listing S1:** A simple python script, illustrating how to use the GAP framework and `quippy` to predict local energies. As command-line arguments, this script takes (1) the coordinate file to be studied (extended `xyz` format), and (2) the GAP potential parameter file (`xml` format).

```python
#!/Users/vld24/anaconda/bin/python2

from quippy import *
import sys

a = Atoms(sys.argv[1])

pot = Potential('IP GAP', param_filename=sys.argv[2])

a.set_calculator(pot)
a.set_cutoff(5.0)

a.get_potential_energies()

e = farray(0.0)
pot.calc(a, energy=e, args_str='local_gap_variance')

a.write('LOCAL_E_and_error_with_'+sys.argv[2]+'_'+sys.argv[1])
```

The information is included in a copy of the structure file, which is now amended by two additional columns in the extended `XYZ` format. The following is an example:

**Listing S2:** Example file header showing how the local energies and GAP variance are written by the script above. In the second line of the file, two new properties are listed (highlighted in green and cyan, respectively). For each atom (starting in the third line), these quantities are then given: note that for the local energy, these values are referenced to that of an atom in diamond-type $c$-Si (evaluated with the same GAP; $\varepsilon(c\text{-Si}) = -163.17763895$ eV), and that the errors are printed as variance (hence the square root must be taken to obtain a quantity in eV).

```
512

pbc="T T T" Lattice="21.90657164      0.00000000      0.00000000
   -0.01012491    21.96387922      0.00000000      0.20206325
   -0.08542605    21.79612182" Properties=species:S:1:pos:R:3:Z:I:
1:local_energy:R:1:local_gap_variance:R:1

Si    7.41226937    21.40807189    17.78143342   14   -163.08144625
0.00001970

Si    ...
```

Visualization was done using the freely available `OVITO` software (https://ovito.org/), which offers a convenient option to color-code atoms according to atomic properties.[S17]

# Supplementary discussion (I): Characterization of *a*-Si networks

To further characterize the structural models that form the basis of this study, we provide key information about the most important ones in **Table S2** as well as additional raw data in **Table S3**. For this analysis, the structures were further relaxed using VASP.[S18,19] This changes the atomic positions only slightly: the GAP model used for melt–quenching is fitted to Perdew–Wang 91 (PW91) data, whereas the further relaxation here is done with the more widely used Perdew–Burke–Ernzerhof (PBE) functional[S20] for wider comparability. It does not affect the topological properties (connectivity, rings) of the networks.

The analysis presented here shows that our slowest-quenched structure outperforms the WWW model in terms of energies while having the same density and average coordination number. Furthermore, the simulated structure factor, $S(q)$, can be used as an additional measure for the quality of the structures because it can be compared to experimental data (**Figure S1**).

All structural models are provided as separate Supporting Information file (a `ZIP` archive containing structural data in extended `XYZ` and VASP `POSCAR` format).

**Table S2:** Information about relevant *a*-Si models, including relative DFT-PBE energies.

|  | $\rho_0$ (g cm$^{-3}$) | Coordination statistics (%) | | | $N_{avg}$ | $\Delta E$ (eV at.$^{-1}$) |
| --- | --- | --- | --- | --- | --- | --- |
|  |  | $N = 3$ | $N = 4$ | $N = 5$ |  |  |
| GAP (10$^{14}$ K/s) | 2.28 | 1.4 | 93.4 | 5.3 | 4.040 | +0.087 |
| GAP (10$^{13}$ K/s) | 2.26 | 1.0 | 96.5 | 2.5 | 4.016 | +0.037 |
| GAP (10$^{12}$ K/s) | 2.25 | 0.4 | 98.4 | 1.2 | 4.007 | +0.008 |
| GAP (10$^{11}$ K/s) | 2.26 | 0.6 | 98.4 | 1.0 | 4.004 | **–0.022** |
| GAP (10$^{10}$ K/s) | 2.26 | 0.6 | 98.8 | 0.6 | 4.000 | **–0.023** |
| WWW | 2.24 | – | 100 | – | 4.000 | 0 (reference) |

**Table S3:** Total energies (raw values); median and maximum force magnitude on atoms for the different models, showing that all structures are well relaxed.

|  | $E$ (eV cell$^{-1}$) | $|\mathbf{F}|_{med}$ (eV Å$^{-1}$) | $|\mathbf{F}|_{max}$ (eV Å$^{-1}$) |
| --- | --- | --- | --- |
| GAP (10$^{14}$ K/s) | –2656.178043 | 0.002 | 0.019 |
| GAP (10$^{13}$ K/s) | –2681.359737 | 0.002 | 0.009 |
| GAP (10$^{12}$ K/s) | –2696.507605 | 0.003 | 0.010 |
| GAP (10$^{11}$ K/s) | –2711.581411 | 0.003 | 0.008 |
| GAP (10$^{10}$ K/s) | –2712.058318 | 0.002 | 0.007 |
| WWW | –2700.479984 | 0.002 | 0.008 |

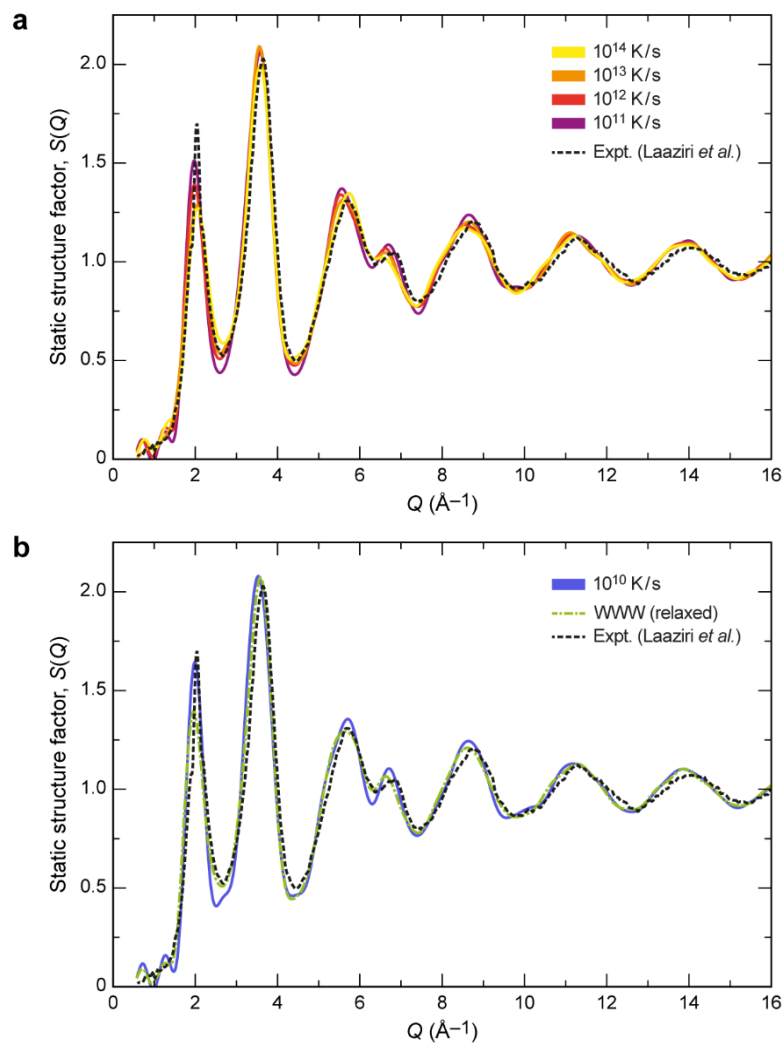

**Figure S1:** Static structure factors as a further means of validation for the different structure models. (a) Evolution of structure factors with decreasing quench rate, similar to [S6] but now for the structures generated independently in the present work. Experimental data from Laaziri *et al.* (dashed black line; Ref. [S21]). (b) Structure factors for the $10^{10}$ K/s quenched structure (blue solid line) and the relaxed WWW network of similar size (dash-dotted green line), again showing experimental data from [S21] for comparison (dashed black line).

# Supplementary discussion (II): Uncertainty quantification

The analysis of local energies can be further supported by the fact that Gaussian process regression makes it possible to quantify the expected error for any prediction. In other words, uncertainty quantification (an extremely important feature of ML models) is directly included in the GAP framework.[S1]

Recall that we obtain the atomic energy by summing up, over $N$ reference configurations ($j = 1, ..., N$), all pairs of fit coefficients $\alpha_j$ and kernel values $K_{ij}$:

$$\varepsilon_i = \sum_j \alpha_j K(\mathbf{q}_i, \mathbf{q}_j) \equiv \sum_j \alpha_j K_{ij},$$

This sum can be written as a dot product if we collect all entries for $\alpha_j$ and $K_{ij}$ (for $j = 1, ..., N$) into appropriate vectors, $\boldsymbol{\alpha}$ and $\mathbf{k}^{(i)}$:

$$\varepsilon_i = \boldsymbol{\alpha} \cdot \mathbf{k}^{(i)} \text{ with } \boldsymbol{\alpha} = \begin{pmatrix} \alpha_1 \\ \vdots \\ \alpha_N \end{pmatrix} \text{ and } \mathbf{k}^{(i)} = \begin{pmatrix} K_{i,1} \\ \vdots \\ K_{i,N} \end{pmatrix}.$$

We can then obtain the variance of the prediction (for the $i$-th atom) as follows:[S1]

$$\sigma_i^2 = K_{ii} - \left(\mathbf{k}^{(i)}\right)^{\mathrm{T}} (\mathbf{K} + \sigma_e \mathbf{I})^{-1} \mathbf{k}^{(i)}$$

where $\sigma_e$ is a small regularization value (on the order of $10^{-4}$ eV; [S1]). By taking its square root, we obtain an error measure with a dimension of energy:[S1]

$$\sigma_i = \sqrt{K_{ii} - (\mathbf{k}^{(i)})^{\mathrm{T}} (\mathbf{K} + \sigma_e \mathbf{I})^{-1} \mathbf{k}^{(i)}}$$

We refer to this as a "prediction error", which we evaluated and plotted in **Figure S2** for the two structures characterized in Figures 3b ($10^{14}$ K/s quench) and Figure 3c ($10^{11}$ K/s quench). Additionally, numerical values for the minimum, median, and maximum prediction errors are provided in **Table S4**.

**Table S4:** Prediction errors for GAP atomic energies (in eV per atom) for the two different structures characterized in Figures 3b–c.

|  | Fast-quenched $a$-Si ($10^{14}$ K/s) | Slow-quenched $a$-Si ($10^{11}$ K/s) |
| --- | --- | --- |
| Maximum ($\sigma_i$) | 0.0193 | 0.0109 |
| Median ($\sigma_i$) | 0.0048 | 0.0037 |
| Minimum ($\sigma_i$) | 0.0022 | 0.0018 |

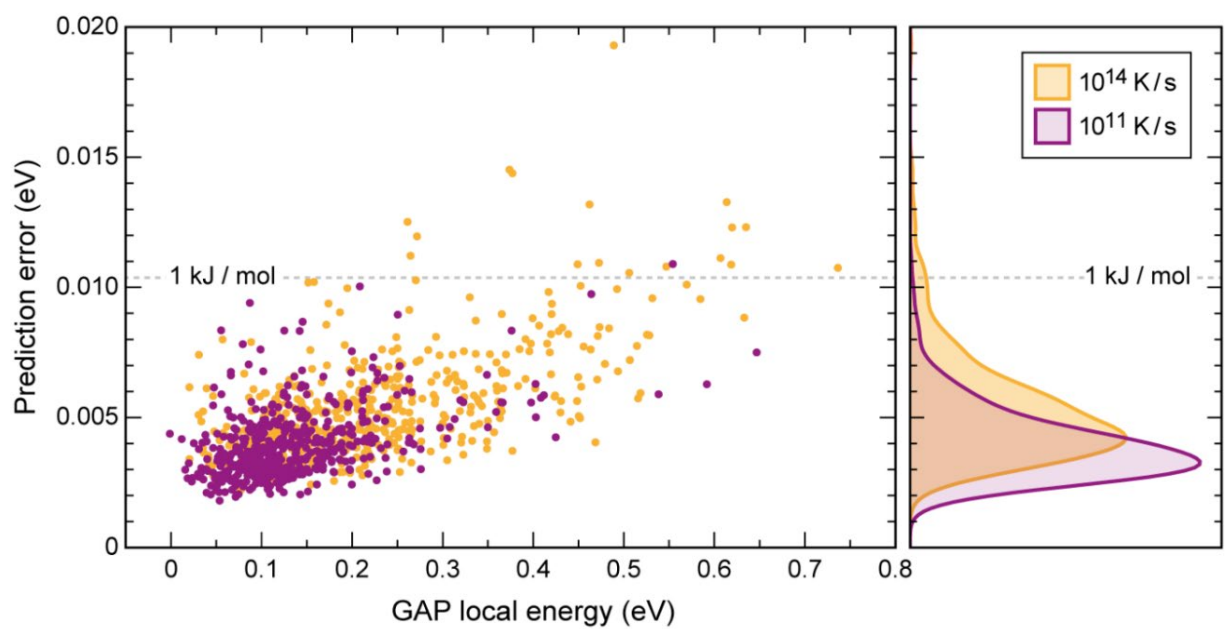

**Figure S2:** Uncertainty quantification by means of the predicted error for GAP local energies. The left-hand side shows all individual values plotted versus the corresponding atom's local energy; the right-hand side shows a kernel density estimate for a projection onto the *y*-axis. A line at 1 kJ / mol is given as an arbitrary reference and shows that the overwhelming majority of data points are below this value, typically taken to be the limit of "chemical accuracy".

## Supplementary references

[S1] A. P. Bartók, J. R. Kermode, N. Bernstein, G. Csányi, *Phys. Rev. X* **2018**, in press, preprint at arXiv:1805.01568.

[S2] S. Plimpton, *J. Comput. Phys.* **1995**, *117*, 1–19.

[S3] M. Parrinello, A. Rahman, *J. Appl. Phys.* **1981**, *52*, 7182–7190.

[S4] G. J. Martyna, D. J. Tobias, M. L. Klein, *J. Chem. Phys.* **1994**, *101*, 4177–4189.

[S5] W. Shinoda, M. Shiga, M. Mikami, *Phys. Rev. B* **2004**, *69*, 134103.

[S6] V. L. Deringer, N. Bernstein, A. P. Bartók, R. N. Kerber, M. J. Cliffe, L. E. Marbella, C. P. Grey, S. R. Elliott, G. Csányi, *J. Phys. Chem. Lett.* **2018**, *9*, 2879–2885.

[S7] A. P. Bartók, M. C. Payne, R. Kondor, G. Csányi, *Phys. Rev. Lett.* **2010**, *104*, 136403.

[S8] A. P. Bartók, R. Kondor, G. Csányi, *Phys. Rev. B* **2013**, *87*, 184115.

[S9] A. P. Bartók, G. Csányi, *Int. J. Quantum Chem.* **2015**, *115*, 1051–1057.

[S10] V. L. Deringer, G. Csányi, *Phys. Rev. B* **2017**, *95*, 094203.

[S11] S. Fujikake, V. L. Deringer, T. H. Lee, M. Krynski, S. R. Elliott, G. Csányi, *J. Chem. Phys.* **2018**, *148*, 241714.

[S12] J. Behler, *Angew. Chem. Int. Ed.* **2017**, *56*, 12828–12840.

[S13] N. Artrith, A. Urban, *Comput. Mater. Sci.* **2016**, *114*, 135–150.

[S14] E. L. Kolsbjerg, A. A. Peterson, B. Hammer, *Phys. Rev. B* **2018**, *97*, 195424.

[S15] S. Maintz, V. L. Deringer, A. L. Tchougréeff, R. Dronskowski, *J. Comput. Chem.* **2013**, *34*, 2557–2567.

[S16] S. Maintz, V. L. Deringer, A. L. Tchougréeff, R. Dronskowski, *J. Comput. Chem.* **2016**, *37*, 1030–1035.

[S17] A. Stukowski, *Model. Simul. Mater. Sci. Eng.* **2010**, *18*, 015012.

[S18] G. Kresse, J. Furthmüller, *Phys. Rev. B* **1996**, *54*, 11169–11186.

[S19] G. Kresse, D. Joubert, *Phys. Rev. B* **1999**, *59*, 1758–1775.

[S20] J. P. Perdew, K. Burke, M. Ernzerhof, *Phys. Rev. Lett.* **1996**, *77*, 3865–3868.

[S21] K. Laaziri, S. Kycia, S. Roorda, M. Chicoine, J. L. Robertson, J. Wang, S. C. Moss, *Phys. Rev. B* **1999**, *60*, 520–533.